\documentstyle[aps,pre,multicol]{revtex}
\draft

\begin{document}
\begin{title}
{\bf Discrete Nonlinear Schr{\"o}dinger Breathers in a Phonon Bath}
\end{title}

\author{K.~{\O}. Rasmussen, S. Aubry\cite{serge}, A.~R. Bishop and G.~P. Tsironis\cite{george} }
\address{Theoretical Division and Center for Nonlinear Studies, Los Alamos National Laboratory, Los Alamos,
NM 87545}
\date{\today}
\maketitle

\begin{abstract}
We study the dynamics of the discrete nonlinear Schr{\"o}dinger lattice initialized such that a 
very long transitory period of time in which
standard Boltzmann statistics is insufficient is reached. 
Our study of
the nonlinear system locked in this {\em non-Gibbsian} state focuses on the
dynamics of discrete breathers (also called intrinsic localized modes).
It is found that part of the energy spontaneously
condenses into several discrete breathers. Although these discrete breathers are
extremely long lived, their total number is found to decrease as the evolution progresses. 
Even though the
total number of discrete breathers decreases we report the surprising observation that the energy content
in the discrete breather population increases. We interpret these observations
in the perspective of discrete breather
creation and annihilation and find that the death of a discrete breather cause effective energy transfer
to a spatially nearby 
discrete breather. It is found that the concepts of a multi-frequency discrete breather and of
internal modes is crucial for this
process. Finally, we find that the existence of a discrete breather
tends to soften the lattice in its immediate neighborhood, resulting in high amplitude thermal fluctuation
close to an existing discrete breather. This in turn nucleates 
discrete breather creation close to a already existing
discrete breather.

\end{abstract}


\section{INTRODUCTION}
The concept of discrete breathers has in recent years become more and more
central \cite{CP,ST,DP,MA,RHTG,CBG} in the investigations of the dynamical properties of nonlinear lattice 
systems. Much research has been devoted to the study of the existence, stability, mobility etc. 
of the discrete breathers, also referred to as "intrinsic localized modes", and already 
several review articles \cite{Aubry,Flach,Tsironis} have been devoted to the subject of 
the dynamics of nonlinear lattices from the perspective of the breather. The main body of this 
work has been devoted to the study of the properties of discrete breathers in homogeneous 
lattices and numerical algorithms
\cite{marin} has been developed allowing accurate and extensive studies of the 
dynamics of pure breather excitations. Recently, their mobility \cite{tsironis,TC} has been 
studied and related to the existence of internal modes of the discrete breathers, which has 
also lead to an appreciation of other properties resulting from 
the internal structures discrete breather can 
posses.

Fewer studies 
have been devoted to discrete breathers in  more realistic environments, such as thermalized systems\cite{CRE},
and to their interaction with other discrete breathers and 
phonons\cite{CAF,cai}, etc. 
In the present work we will 
study breathers or breather-like excitations that are spontaneously created and thereafter 
exists in a realistic noisy environment.
The spontaneous emergence of breathers has previously been observed in contexts driven
by thermal shocks, \cite{ATetal} where the environment is almost void any thermal fluctuations.

We  perform this study in the framework of a one-dimensional discrete nonintegrable nonlinear Schr{\"o}dinger 
(DNLS) equation. The existence and stability and several other aspects of breathers in the DNLS 
system have been clarified using the concept of "anti-integrability" \cite{MAG}, and Melnikov analysis
\cite{hennig}. The importance of the DNLS system, as one of the most widely studied discrete
nonlinear systems, stems not only from its applicability in diverse physical situations but also
from its simple, yet rich, mathematical structure. It is worth mentioning that the DNLS 
exhibits a richness that this system in its integrable version does not, either in the discrete
version (Ablowitz-Ladik discretization \cite{AL}) or its continuum limit.

The thermalization of the DNLS systems has been studied previously \cite{nogood,CRCG} and 
these studies have shown that the dynamics exhibits two very different 
characteristics depending on the initial energy density $h={H}/N$ and norm density 
$n={\cal N}/N$ (for definition of Hamiltonian ${H}$ and norm ${\cal N}$ see, Eq. (\ref{eq3})
and below Eq. (\ref{eq3}), respectively), where $N$ is the system size.
At low densities (exact relationship is given in \cite{CRCG}) a thermalized state appears 
after a relatively short time, where all the correlations can be obtained 
from the partition function
\begin{equation} 
Z(\beta)={\mbox Tr} \left (\exp\left [-\beta({ H} +\mu {\cal N})\right ] \right ).
\end{equation}
In the thermalized state the discrete
breathers do not occur with long lifetimes and are not easily distinguishable from the 
phonon background. 

Contrary at high densities a {\em non-Gibbsian} regime where the a statistical 
mechanics description requires {\em formal} introduction of ``negative temperatures'' emerge.
In this regime  breather-like excitations spontaneously appear in the dynamics. This is 
much in accordance with 
the concepts known in the statistical mechanics description of vortex structures 
in plasma and hydrodynamics applications \cite{oldstuff}. A complete discussion of the 
statistical mechanics results is given in \cite{CRCG}. Since, the existence of 
long transitory regimes is known several other contexts the above described concept may be 
thought of as somewhat generic to extended nonlinear systems. Therefore the focus of this paper 
is the dynamics in this non-Gibbsian regime.

The structure of the paper is a follows. In the next section, we describe the numerical
simulations we have performed on the system. We will 
describe in detail the 
observations, focusing on the breather behavior. In Sec. III we give our interpretations
of the observation reported in Sec. II. Finally, Section IV presents our
conclusions.

\section{MOLECULAR DYNAMICS RESULTS}
\label{II}

We are studying the one-dimensional discrete nonlinear Schr{\"o}dinger equation (DNLS) 
\begin{equation}
i\dot \psi_n+(\psi_{n+1}+\psi_{n-1})+a|\psi_n|^2\psi_n=0,
\label{eq1}
\end{equation}
where the overdot denotes the time derivative, $n$ is a site index, and $a$ is a real tunable coefficient to 
the nonlinear term.
Equation (\ref{eq1}) can be written in Hamiltonian form as
\begin{equation}
i\dot \psi_n=\frac{\partial H}{\partial \psi^*_n}
\label{eq2}
\end{equation}
with canonical conjugated variables $\{ \psi_n,i\psi^*_n\}$ and Hamiltonian
\begin{eqnarray}
H(\{\psi_n,&&i\psi_n^*\})=\nonumber \\&& \sum_n \left ( -(\psi_n\psi_{n+1}^*+\psi_n^*\psi_{n+1})
-\frac{a}{2}|\psi_n|^4 \right ).
\label{eq3}
\end{eqnarray}
In addition to the Hamiltonian $H$, the dynamics conserves the norm ${\cal N}=\sum_n |\psi_n|^2$. These
conserved quantities were monitored frequently in our molecular dynamics simulations to insure accuracy of the 
4'th order Runge-Kutta scheme.

The system is initialized by assigning each site with a random value, $x$, with the 
distribution, $p(x)$
\begin{equation}
p(x)=\frac{k}{\pi}\frac{1}{x^2+k^2}.
\label{eq4}
\end{equation}

The Lorentzian distribution is chosen in order to have more than exponentially small
probabilities for sites being assigned large values which leads the breather-like excitations 
to appear almost immediately. Several different initial conditions like, uniform distributions,
single phonon modes, etc. has been tried and as long as the energy density and norm densities 
are in the non-Gibbsian regime (see, \cite{CRCG})  the initial configuration is of no importance 
in the sense that the breather spontaneously appear and show the behavior we are about to describe. 
The initial configurations does however influence the timescale involved in reaching the breather 
regime, which is the reason we choose the Lorentzian (\ref{eq4}).

As an example of the molecular dynamics simulations we show 
in Fig.1 the evolution of the energy density $E_n$. Here, we clearly see
\begin{figure}
\vspace*{270pt}
\hspace{-50pt}
\includegraphics{`zcat}
\vspace*{-70pt}
\caption{Evolution of local energy $E_n$ along the chain (total length 2500 sites). 
The horizontal axis indicates the position along 
the chain and the vertical axis corresponds to time (evolving downwards). The blue color indicates that 
the local energy is at its background level while the more red indicate energies higher than
background level. $a=10$ and the initial state was taken from Eq.(\ref{eq4}) with $k=0.1$}
\label{fig1}
\end{figure}
that high amplitude excitations spontaneously appear out of the 
fluctuations. That these excitations 
are breathers is easily verified by frequency analyzing the dynamics of these particular sites. 
\begin{figure}
\vspace*{270pt}
\hspace{-20pt}
\includegraphics{`zcat}
\vspace*{-70pt}
\caption{The fraction of the total energy $E_0$ trapped at the breathers $E_b$, versus time. Here a 
breather is (dashed line) defined by an amplitude higher than $1.2$ and (solid line) higher than $1.3$}
\label{fig2}
\end{figure}
It is seen that some of these breathers are extremely long lived, while others vanish. It is therefore tempting to 
conclude that eventually all breathers will vanish from the system causing it to relax into an 
equipartitioned state. However, if we monitor the evolution of the part of the total energy $E_0$ which 
is contained within the breather, we find that the fraction of the energy trapped in the breathers is increasing. 
An example obtained for a $N=2500$ site system is shown in Fig.2.
In Fig.2 the breather has been defined (dashed line) by an amplitude higher than $1.2$ and 
(solid line) higher than $1.3$.
Even though the number of breathers is obviously decreasing, 
the energy content in the breathers is increasing, causing an effective cooling of the 
phonon bath. A closer scrutiny of 
the curves (particular the dashed curve) shows 
that the energy content in the breathers remains constant for a long 
period of time and then rather suddenly increases. Afterwards it stays constant again 
until the sudden increase repeats itself.
This indicates that the increase is related to special events and thus not a continuous 
process. This was tested by inserting an exact breather into a previously thermalized 
bath with a given temperature $T$\cite{CRCG} and then monitoring the 
energy content in the breather. The experiment was performed for a representative set of bath 
temperatures and breather amplitudes (frequencies) and the breather was never observed to 
gain energy from the bath; rather the energy content always remained (apart from small fluctuations)  
constant for thousand of breathing periods unless the discrete breather was 
destroyed at an early stage by the thermal fluctuation. The spontaneous destruction happened
primarily for small amplitude breathers and we will address this phenomenon again later 
in this paper.
\begin{figure}
\vspace*{270pt}
\hspace{-20pt}
\includegraphics{`zcat}
\vspace*{-70pt}
\caption{Expanded scale showing a single frequency breather spontaneous splitting  into 
several smaller amplitude excitations. }
\label{fig3}
\end{figure}

We take the above observations as a demonstration that the 
breathers can not pump energy directly from a thermalized background.
 A direct indication of the energy gain in the breathers being related to discrete
events is found in Fig.1. Here, we clearly see that the increase of energy content in the right most 
breather occurs just after the smaller breather to the left vanishes. We have observed this 
to be the typical scenario. Thus there is a tendency for large amplitude 
breathers to absorb some of the energy released once a spatially close by breather dies. 
The tendency of the energy to accumulate in breather exitations has previously been obeserved \cite{bang}.

In Fig.3 we show a blow-up of the 
time window where a breather vanishes and it is seen that the breather splits up into several small amplitude 
but localized 
excitations which propagate in apparently random directions.

Figure 4 displays the evolution of a center site and the two nearest-neighbor sites for a 
breather that is about to break up into several small amplitude excitations. As the breather approaches the 
break-up point the amplitude at the neighboring site to the left increases significantly, and 
additionally the oscillation at the center site introduce a new frequency besides the 
breathing frequency. 
\begin{figure}
\vspace*{290pt}
\hspace{-30pt}
\includegraphics{`zcat}
\vspace*{-90pt}
\caption{The evolution of $|\psi_n|$ just before the breather breaks up. Shown is the center 
site $n=0$ (solid line), the neighboring site to the left $n=-1$ (dashed line) where the new breather 
forms, and the neighboring site to the right  $n=1$ (dashed-dotted line) where the evolution is essentially 
unaffected}
\label{fig4}
\end{figure}

If we Fourier analyze the evolution shown in Fig. 4 (see in Fig.5) 
we find that the frequency of the oscillation at the neighboring site to the left increases until 
a certain point amplitude (frequency). It is 
noteworthy that the frequency is rather far outside the linear phonon band. Additionally 
the center oscillation is also affected and the frequency of the neighboring site 
is apparent in the spectrum of the center site. 

This scenario indicates that the breather becomes unstable because high amplitude 
is accumulating at the neighboring site, causing the breather to transform into 
a multi-breather state.  The splitting of the breather can be analyses in terms of the 
stability of this multi-breather state.

\begin{figure}
\vspace*{300pt}
\hspace{-50pt}
\includegraphics{`zcat}
\vspace*{-90pt}
\caption{ Frequency spectrum of the evolution shown in Fig.4 for the center site $n=0$ (upper panel) 
and for the neighboring site to the left $n=-1$ (lower panel)} 
\label{fig5}
\end{figure}

\section{INTERPRETATION OF MOLECULAR DYNAMICS RESULTS}
In this section we will present an interpretation of the observed phenomena in terms of 
breather dynamics. The basic elements are that initial condition prevents
the system from directly relaxing into a Gibbsian thermalized state. Instead a very long ($> 10^6$ time 
units) transitory state is reached, where the system behavior is dominated by spontaneously 
created breathers. Although the number of breathers
is decreasing, the portion of the energy trapped in these excitations is 
increasing. We interpret this as follows. Due to the increased amplitude of the fluctuations 
close to an existing breather (shown in subsection \ref{sub2}) a two- (or more) frequency 
breather is created. An instability due to an internal mode in these more complex excitations 
can (rather than the initial breather)
 provide several (shown in subsection \ref{sub1}) paths for the initial 
breather excitation to split into lower amplitude 
excitations capable of propagation. The generated propagating excitations can
interact with other existing breathers and transfer their (or part of their) energy 
to stationary high amplitude breathers (addressed in subsection \ref{sub3}) 
producing transfer of energy from the fluctuating phonon bath to the high amplitude breathers.

\subsection{TWO FREQUENCY BREATHERS} 
\label{sub1}
It has already been demonstrated in Figs. \ref{fig4} and \ref{fig5} that an 
excitation with more that one frequency can form at the site of an already 
existing single frequency breather. In order to study the splitting process 
shown in Fig. \ref{fig3} we investigate the stability of a two-frequency 
breather.

The linear stability of multi-breathers in the framework of the DNLS equation has 
already been discussed \cite{MAG} in the spirit of the anti-continuous limit \cite{MA}.
That is starting from the decoupled system and numerically continuing the trivial 
breather solution into the regime of non zero coupling. Here we shall briefly study the changing 
stability as ratio between the two breathing frequencies,
$\omega_o$ and $\omega_b$, changes. Specifically we shall study the stability of the 
two frequency breather keeping $\omega_b$ fixed and changing $\omega_0$, ($\omega_0 <  \omega_b$).

The linear stability is determined by the spectrum\cite{MAG} of the Floquet 
${\mathbf F}$ operator (defined in Eq.(\ref{Teq2})), which we calculate and 
diagonalize numerically. Generally, linear stability is assured so long as all 
eigenvalues remain on the unit circle (due to the Floquet operator
being symplectic). For the the two-frequency breather 
four eigenvalues will always be unity $\lambda=1$ due to time reversibility \cite{MAG}.

The stability of a two-frequency breathers generally follows the following  
pattern for fixed frequency $\omega_b$ (and nonlinearity $a$). The two-frequency 
breather is unstable if $\omega_o$ is too small, but becomes stable as 
$\omega_o$ is increased. Further increasing $\omega_o$ leads, at $\omega_0=\omega_b-2$ to
the localized two-frequency breather bifurcating into a {\em phonon breather} which 
has a non vanishing phonon tail. Investigating the the stability in more detail 
we therefore choose ($\omega_b=14$) (which is in the range of the breathers observed
in the full system dynamics) and choose $\omega_o=3.4$ where the two-frequency breather is stable. In 
Fig. 6 we show the eigenvalue of the Floquet matrix as the frequency $\omega_o$ is decreased.

\begin{figure}
\vspace*{300pt}
\hspace{-45pt}
\includegraphics{`zcat}
\vspace*{-90pt}
\caption{The Floquet spectrum $\lambda =|\lambda|\exp(i\theta)$ versus
versus frequency $\omega_o$. Upper panel shows $|\lambda|$ vs. $\omega_o$ and 
lower panel shows $\theta$ vs. $\omega_o$.($\omega_b=14$)}
\end{figure}

At $\omega_o=3.4$ all the eigenvalues are in the phonon band (strictly speaking the 
frequency $\omega_{\mu}=\frac{\theta}{2\pi}\omega_b$ ($\lambda=\exp(\pm i\theta)$) of the corresponding 
eigenmode is in the phonon band) so all eigenmodes are extended. However, as $\omega_o$
is decreased one pair of eigenvalues bifurcates off the $k=0$ phonon band edge (at $\omega_o \simeq 3.2$),
and creates a localized eigenmode. These eigenvalues collide at unity when $\omega_o \simeq 2.9$ and
move out on the real axis as $\omega_o$ is decreased, further causing the two-frequency breather 
to become unstable. If we increase $\omega_o$, no localized modes appear 
and the breather consequently remains stable. However when $\omega_o=\omega_b-2$
(as can be deduced from Eq.(29) of Ref. \CITE{MAG}) the two-frequency breather
bifurcates to a phononbreather shown in Fig.7. The phonon breathers is also linearly 
stable but because of the nonvanishing tail the energy content is not finite making this 
object irrelevant for our case.

\begin{figure}
\vspace*{300pt}
\hspace{-50pt}
\includegraphics{`zcat}
\vspace*{-90pt}
\caption{Two-frequency breather versus $\omega_o$. A bifurcation seen at $\omega_o=12$. $\omega_b=14$}
\label{fig7}
\end{figure}

It was already demonstrated in  Ref. \CITE{MAG} that localized mode bifurcating off the 
linear band corresponds to a pinning mode of the weakest localized part of the two-frequency breather. Also 
in Ref. \CITE{tsironis} it was demonstrated for a single-site breather that a perturbation along the 
pinning mode can lead to mobility of the breather. In our situation this implies that a perturbation along 
the localized mode could lead the two-frequency breather to break apart into its two components. The presence of
a thermalized bath around the breather could lead to a perturbation of the localized mode and consequently 
a spontaneous splitting. To illustrate this, we have injected a two-frequency breather 
($\omega_b=14$, and $\omega_o=3.5$; thus a breather with a very weak localized mode) into 
a bath thermalized at the (low) temperature $\beta=(k_BT)^{-1}=1000$ (see Ref. \CITE{CRCG} for 
details about thermalizing the system at a predescribed temperature). The result is seen 
in Fig.8 where the energy density is plotted versus time. As expected, the presence 
of the thermal bath causes the two-frequency breather to split into its to components. The splitting 
was tested not to occur in the absence of the localized mode.
\begin{figure}
\vspace*{300pt}
\hspace{-50pt}
\includegraphics{`zcat}
\vspace*{-90pt}
\caption{Energy density of the evolution of a two-frequency breather 
versus time ($\omega_b=14$, and $\omega_o=3.5$) subjected to a thermalized bath.}
\label{fig8}
\end{figure}

We believe this mechanism is responsible for the thermally induced splitting that is observed in the 
full dynamics, of which an example is detailed in Fig. \ref{fig3}.

\subsection{TWO FREQUENCY BREATHER GENERATION}
\label{sub2}

In this section we address question of why the presence of a single frequency breather allows creation of a nearby breather 
which eventually leads to the creation of  breathers.
Considering the statistics of the phonon bath around a breather, we assume that it is 
acceptable to treat the small fluctuations in a linear framework and therefore study 
the linearized system around a periodic breather solution $\{ \psi_n^{(0)}(t) \}$. That is we 
have $\psi_n(t)=\psi_n^{(0)}(t)+\epsilon_n(t)$. Decomposing into real and 
real and imaginary parts, we write $\psi_n^{(0)}=x_n+iy_n$, and 
$\epsilon_n=\xi_n+i\eta_n$, where $x_n$, $y_n$, $\xi_n$, and $\eta_n$ are 
real functions. Substituting into Eq.(\ref{eq1}) and linearizing, we obtain 

\begin{eqnarray}
\dot \xi_n=&-&(\eta_{n+1}+\eta_{n-1})\nonumber \\&+&a((x_n^2+y_n^2)\eta_n-2(x_n\xi_n+y_n\eta_n)y_n)\\
\dot \eta_n=&-&(\xi_{n+1}+\xi_{n-1})\nonumber \\&+&a((x_n^2+y_n^2)\xi_n-2(x_n\xi_n+y_n\eta_n)x_n).
\label{Teq1}
\end{eqnarray}
Then the linear modes around the breather are 
found by diagonalizing the Floquet operator, ${\mathbf F}$, defined as

\begin{eqnarray}
X(t_b)= {\mathbf F} X(0)
\label{Teq2}
\end{eqnarray}
with 
\begin{equation}
X(t) = \left(\begin{array}{c}\{\xi_n(t)\}  \\
\{{\eta}_{n}(t)\} \end{array} \right).
\end{equation}

Here, $\mathbf{F}$ exhibits a set of conjugate pairs of eigenvalue $\exp(\pm2\pi i \omega_{\mu}/\omega_b)$
and the corresponding {\em normalized} eigenmodes $X^{\mu}(0)=\{\xi_n^{\mu}(0),\eta_n^{\mu}(0) \}$
and ${X^{\mu}}^*(0)$. These eigenmodes will fulfill the Bloch condition 
$X^{\mu}(t)=\exp(i\omega_{\mu} t)\chi_{\mu}(t)$, where $\chi_{\mu}$ is periodic with the period
$t_b$ of the breather. 

It is known \cite{CAF} that as we are dealing with a one channel scattering the scattering of 
phonons through the breather is elastic, so that the breathers behave as a {\em static impurity} 
at least in the limit of low amplitude phonons.
In this case it appears justified that,
in thermal equilibrium at temperature $T=(k_B \beta)^{-1}$, the complex amplitude $\lambda_{\mu}$
of each eigenmode follows standard Boltzmann statistics $\exp(-\beta \omega_{\mu}|\lambda_{\mu}|^2/2)/C_0$
($C_0$ is a normalizing factor). This will produce the the correct thermodynamic statistics
far from the breather, where the eigenmodes are plane waves with frequency $\omega_{\mu}$ and wave vector
$\pm q_{\mu}$. However, this requires $\epsilon^{\mu}_n$ to be expressed in an orthonormal basis
\begin{equation}
\sum_n|\epsilon^{\mu}_n|^2=1,~~~~\mbox{and}~~~~~ \sum_n{\epsilon^{\mu}}^*_n \epsilon^{\mu'}_n=
\delta_{\mu,\mu'}. 
\label{Teq4}
\end{equation}
This condition would be trivially achievable considering a real static impurity. It is more 
speculative in our case, with a time dependent potential. In a small system these 
quantities may dependent on time, for large systems however this time dependence
becomes negligible because the solutions of Eq. (\ref{Teq1}) tend to plane waves at infinity 
and the weight of the breather region thus become negligible. (The problem would persist
in the presence of internal modes but we assume that such 
modes are absent). So the remedy of this problem is to choose the system larger 
enough, in order to avoid significant time dependence.

Now the general solution of the linearized equations (\ref{Teq1}) can be expressed as a 
linear combination of the eigenmodes
\begin{equation}
\epsilon_n=\frac{1}{2}\sum_{\mu}\lambda_{\mu}\epsilon_n^{\mu}+\mbox{cc}.
\label{Teq5}
\end{equation}
The probability distribution of $\epsilon_n$ which is a sum of independent
Gaussian variables, is obviously, from Eq. (\ref{Teq5}), a Gaussian. The variance of 
the Gaussian is 
\begin{equation}
\langle |\epsilon_n|^2 \rangle = \sum_{\mu} \langle |\lambda_{\mu}|^2\rangle|\epsilon^{\mu}_n|^2 =
\sum_{\mu} \frac{1}{\beta \omega_{\mu}}|\epsilon^{\mu}_n|^2.
\label{Teq6}
\end{equation}
In a large system without a breather the linear modes would be plane waves,
and Eq. (\ref{Teq6}) would yield the standard result

\begin{equation}
\langle |\epsilon_n|^2 \rangle = \frac{1}{2\pi\beta}\int_{-\pi}^{\pi} \frac{dq}{\omega (q)},
\label{Teq7}
\end{equation}
where $\omega (q)$ is the phonon dispersion. Thus, in the absence of the breather the 
average size of the fluctuations would be uniform. In the presence 
of a breather we have to calculated $|\epsilon_n|^2$ numerically, diagonalizing the 
Floquet operator. In Fig. 9 we show $|\epsilon_n|^2$ versus lattice site $n$ and breather
frequency $\omega_b$. Surprisingly, we see that the averaged fluctuation at the sites
close to the breather is significantly larger than in the rest of the lattice. The breather seems
to introduce a softening at the neighboring sites. The possibility of large fluctuations 
close to the breather sites indicates that there is an increased probability to create a new 
breather close to an existing breather, which explains the formation of 
multi-frequency breathers seen in the simulations reported in this paper. Further we see from 
Fig. 9 that, as the breather frequency $\omega_b$ approaches the edge of the linear band, 
the fluctuations at the breather center increase dramatically, indicating that broad, 
low amplitude breathers have a higher probability of being destroyed by the thermal fluctuations
than high amplitude breathers.
\begin{figure}
\vspace*{300pt}
\hspace{-30pt}
\includegraphics{`zcat}
\vspace*{-90pt}
\caption{Variance $\langle |\epsilon_n|^2 \rangle$ of thermal fluctuations, versus breather frequency
$\omega_b$. Breather center is at $n=31$.}
\end{figure}
	
\subsection{ENERGY ACCUMULATION IN BREATHERS}
\label{sub3}

In the preceding two subsections we show how the presence of a single frequency breather
makes it probable that a new breather is generated nearby by the thermal fluctuations, and we 
show that the multi-frequency breathers created in this manner are likely to split 
due to thermal excitation of their internal modes. We are still missing the final feature, namely 
the absorption of the generated small amplitude breathers by  neighboring 
high amplitude breathers. We now address this.

As a demonstration that this is indeed possible, we show an example in Fig. \ref{fig10}.
\begin{figure}
\vspace*{300pt}
\hspace{-30pt}
\includegraphics{`zcat}
\vspace*{-90pt}
\caption{Collision of a propagating small amplitude breather with a pinned large
amplitude breather with $\omega_b=7.5$. Almost total absorption is observed. $a=10$}
\label{fig10}
\end{figure}
Figure \ref{fig10} shows the collision of a small amplitude wave initialized as
\begin{equation}
\psi_n=A\mbox{sech}(Bn)\exp(iCn),
\label{finaleq}
\end{equation}
where $A$, $B$ and $C$ are tunable parameters. (In Fig. \ref{fig10} $A=0.2$ $B=0.3$ and 
$C=0.4$ was chosen). The particular values of these parameters has no significance
for the phenomenon observed: several choices were tested and absorption
always occurred to some degree. The reason for the wavepacket  not to move 
at constant velocity is that, due to the large nonlinearity ($a=10$), the 
localized excitations encounters a rather large barrier \cite{cai&co} when 
translating from site to site.

There is one subtlety to the phenomenon of absorption we demonstrate in 
Fig. \ref{fig10}. Namely, if the large amplitude breather is chosen to be exact, that is 
generated from the anti-continuous limit, and inserted in a non-thermalized system, 
and the small amplitude wavepacket  is launched toward it {\em no} absorption occurs since, 
the wavepacket is fully reflected. On the other hand if the breather $\psi^{s}_n$ upon injection 
in the system is perturbed in  the following manner,
\begin{equation}
\psi_n=\psi^{s}_n\exp(i\gamma n),
\label{finaleqq}
\end{equation}
then the absorption appears as demonstrated above. The power $\gamma$ can be chosen to be
rather large, because the large amplitude breathers is very tightly pinned 
to the lattice \cite{cai&co}.

\section{SUMMARY}
In summary we have found find that initializing the system in a non Gibbsian state with large 
fluctuations spontaneously generates several discrete breathers. The total energy content of these 
breathers increase in time while the actual number of breathers decrease so that the 
system tends to a state of isolated breathers in a cold phonon bath. These breathers become 
more rare but of increasingly larger amplitude and the phonon bath colder as time increases.

Discrete breathers seem to favor larger phonon fluctuations in their immediate neighborhood, which 
makes more probable the generation of a new breather in their close vicinity (using the 
energy of the phonon bath). This new breather may be ejected as a propagating breather
by effect of the phonon noise if this two frequency breather approaches its instability 
threshold. In the example of Fig.9 a stationary single frequency breather is recovered, allowing
the same process to be repeated. This means that the initial breather acts like a 
catalyst for extracting energy from the phonon bath and generating new
propagating breathers.  In other cases, the two-frequency breather is, via the 
interaction with the phonon bath, broken into two (or more) propagating breathers. 
Finally the traveling breathers can be absorbed (or at least partly absorbed) by 
existing breathers. As a result the number of breathers eventually decrease while 
their total energy content increases, leaving the phonon bath increasingly colder.

Work at Los Alamos National Laboratory is performed under the auspice of 
the US DOE.


\end{document}